\documentstyle[11pt,epsf]{article}

\begin{document}

\begin{centering}
{\Large \bf GRS 1915+105: the flux ratio of twin radio clouds as 

              a measure of asymmetry between counter jets}

\end{centering}

\vspace{3mm}

\centerline{ \large A. M. Atoyan$^{1,2}$ and F. A. Aharonian$^{1}$}

\vspace{3mm}

\centerline{\small $^{1}$ 
Max-Planck-Institut f\"{u}r Kernphysik, Postfach 10 39 80,
69029 Heidelberg, Germany}

\centerline{\small $^{2}$  
Yerevan Physics Institute, Alikhanian Brothers 2, 375036 Yerevan, Armenia}
\vspace{3mm}

\begin{abstract}
Resolution of both approaching and receding ejecta in the galactic 
microquasars makes possible to measure the flux ratio $S_{\rm a}/S_{\rm r}$ of
twin ejecta, which contains an important information about the
nature of the jets. We show that the flux ratio 
$S_{\rm a}/S_{\rm r} = 8\pm 1$ observed from GRS 1915+105 during the  
prominent March/April 1994 radio flare can be explained in terms of 
relativistic motion of
discrete radio clouds, if one assumes that the twin ejecta are similar,
but not completely identical, i.e. allowing for some asymmetry between
the plasmoids in their speeds of propagation and/or luminosities.  
The recoil momentum due to asymmetrical ejection of the pair of plasmoids
could be comparable with the momentum accumulated in the inner accretion disk.
We suggest a possible explanation of the observed anticorrelation 
between the X-ray and radio flares as the result of drastic structural
changes in the inner disk caused by production of powerful jets.
The delay between the times of decline of X-rays and 
appearance of strong radio flares is explained by time needed for 
expansion of radio clouds to become optically thin.  

\vspace{1mm}
\noindent
{\em Subject headings:}~~ {stars: individual (GRS 1915+105) -- stars: flare -- 
galaxies: jets
 -- radio continuum:  general -- accretion disks}

\end{abstract}

\section{Introduction}

Recent discovery of the galactic superluminal jet sources GRS 1915+105 
(Mirabel \& Rodriguez 1994, hereafter MR94) and 
GRO J1655-40 (Tingay et al. 1995; Hjellming \& Rupen 1995) provides  
unique possibility  for deep study of the 
phenomenon of relativistic jets observed elsewhere in the Universe. 
These 
sources, called {\it microquasars}, represent, to a large extent, scaled down 
analogs of  AGNs 
(Mirabel \& Rodriguez 1995). 
Being, however,
much closer to us than AGNs, the microquasars enable radio monitoring
of {\it both} approaching and receding relativistic ejecta, and importantly, 
in short timescales.

GRS 1915+105 is the first source of apparent 
superluminal jets where the receding jet has been also detected, and 
time evolution of radio fluxes of both components of the prominent 
19 March 1994 flare has been traced on timescales of days beyond 30
April (MR94). In particular, this has 
allowed Mirabel \& 
Rodriguez  to determine the real speed  $\beta \approx 0.92 $ and the 
angle of propagation $\theta \approx 70^\circ$ of the ejecta. 
At an estimated distance $D = 12.5\, \pm 1.5\,\rm kpc $, the observed 
proper motion of radio sources correspond to apparent velocities 
(in units of speed of light) $v_{\rm a} = 1.25 \pm 0.15$ and 
$v_{\rm r} = 0.65 \pm 0.08$ for the approaching and receding ejecta, 
respectively (MR94). 

Detection of pairs of jets  makes possible
to put forward an important question concerning the origin of the 
radio images observed. 
Namely, interpretation of the ratio of radio fluxes detected from the 
twin jets in GRS 1915+105 actually addresses an 
important issue,
whether the moving radio images are the radio `echos' of 
relativistic shocks propagating in the jets, or they 
correspond to real motion of
discrete plasmoids.

The ratio 
of flux densities, $S(\nu) \propto \nu^{-\alpha}$, measured at 
equal angular separations of the {\it identical} pair of jets from 
the core is  
\begin{equation}
\frac{S_{\rm a}}{S_{\rm r}} = \left( \frac{1\, +\, \beta \, \cos \theta}
{1\, -\, \beta \, \cos \theta} \right)^{ k + \alpha}\; \; \cdot
\end{equation}
The index $k$ is equal to $k=3$, if the fluxes are 
produced in the moving {\it discrete} radio clouds, but $k=2$
for the brightness ratio of continuous {\it stationary } jets 
(e.g. see Lind \& Blandford 1985).
The difference in the index $k$ is due to 
relativistic volume contraction effect (see Bodo \& Ghisellini 1995).
For $\beta=0.92$, $\theta =70^\circ$, 
and measured radio spectral index $\alpha= 0.84 \pm 0.03$ (MR94), 
Eq.(1) results in the flux ratio $S_{\rm a}/S_{\rm r} \simeq 12$ to be 
expected from discrete sources, 
whereas the measured flux ratio $S_{\rm a}/S_{\rm r} = 8 \pm 1 $ (MR94)
is closer to $S_{\rm a}/S_{\rm r} \simeq 6 $ which is formally 
expected in the case of stationary jets. Meanwhile, the apparent 
motion of discrete radio condensations does not support the interpretation 
of the observed flux ratio in terms of emission of stationary jets. 
A possible explanation of this discrepancy between the observed and 
expected (for discrete plasmoids) flux ratios of the jets in GRS 
1915+105 was given by Bodo \& Ghisellini (1995), who suggested that the real 
speed of radioemitting plasma ({\it fluid}) in the jets is  different 
from the 
speed $\beta \approx 0.92$ of radio {\it patterns} attributed to 
propagation of shocks in the fluid.

In this paper  we propose another interpretation of the measured flux
ratio,   in terms of speeds of the ejecta coinciding 
with the speeds  $\beta \sim 0.9$ of radio patterns, assuming that the 
twin ejecta are {\it similar}, but {\it not } absolutely identical.

\section{ Pair of asymmetrical jets}   

When both components of relativistic jets can be detected, then in addition 
to the equation for the angular speed $\mu_{\rm a}$ of the 
approaching jet, two more equations become available, namely, the one for the
angular speed $\mu_{\rm r}\,$ of the receding component, as well as 
the equation for 
the flux ratio of the two components. 
Allowing for the parameters of the counter ejecta to be different, 
the first two equations read:
\begin{equation}
\mu_{\rm a}=\frac{\beta_{\rm a}\sin \theta_{\rm a}}
{1 -\beta_{\rm a}\cos \theta_{\rm a}}\,  
\frac{c}{D} \;\; ,
\end{equation}
\begin{equation}
\mu_{\rm r}=\frac{\beta_{\rm r}\sin \theta_{\rm r}}
{1 +\beta_{\rm r}\cos \theta_{\rm r}} \,
\frac{c}{D} \;\; ,
\end{equation}
where $\beta_{\rm a}$ and $\beta_{\rm r}$ 
are the speeds, $\theta_{\rm a}$ and $\theta_{\rm r}$ are
the angles of propagation of approaching and receding ejecta.
Note that for convenience in Eq.(3) we have substituted 
$\theta_{\rm r}\rightarrow
180^{\circ}-\theta_{\rm r}$.

The equation for the flux ratio $S_{\rm a}/S_{\rm r}$ 
for an asymmetrical pair of ejecta is found, 
using the relation 
$S(\nu) = \delta^{3+\alpha} \, 
S^{\prime}(\nu)$ (e.g., Lind \& Blandford  1985) between the apparent 
and intrinsic energy fluxes of a radio cloud moving with 
Doppler factor 
$\delta = \sqrt{1-\beta^2}/
(1-\beta \cos \theta) $. Then the 
flux ratio at equal {\it intrinsic times} is:
\begin{equation}
\frac{S_{\rm a}}{S_{\rm r}} = 
\left(\frac{\Gamma_{\rm r}}{\Gamma_{\rm a}}\right)^{3+\alpha}
\left(\frac{1+\beta_{\rm r} \cos \theta_{\rm r}}{1-\beta_{\rm a}\cos 
\theta_{\rm a}}
\right)^{3+\alpha} \, \frac{L_{\rm a}^\prime}{L_{\rm r}^{\prime}}\;,
\end{equation} 
where $\Gamma_{\rm a,r}= (1-\beta_{\rm a,r}^2)^{-1/2}$ are the Lorentz factors
of the bulk motion, and $L_{\rm a,r}^\prime$ are the 
intrinsic luminosities of the clouds. 
An asymmetry between the ejecta would imply at least one of the 
following inequalities: 
(a)~ the luminosities of the approaching and receding ejecta
in their rest frame are different, i.e. 
$L_{\rm a}^\prime \neq L_{\rm r}^\prime\,$;
(b)~ the ejecta propagate not strictly in the opposite directions,
i.e. $\theta_{\rm a}\neq \theta_{\rm r}\,$; 
(c)~ the speeds of the ejecta are different,
$\beta_{\rm a} \neq \beta_{\rm r}\,$.
In principle, one may assume also that the spectra 
 of the ejecta  are described by different power-law
indexes $\alpha$. However, since the same  
$\alpha\approx 0.84 $  was observed from both radio 
clouds  of the 19 March 1994 flare, 
we do not consider here such a possibility.

Eq.(4) corresponds to the flux ratio of two radio clouds at equal 
{\it intrinsic times}, $t_{\rm a}^\prime = t_{\rm r}^\prime =t^\prime $.
Meanwhile, the directly measurable quantity is the flux ratio 
at equal {\it angular separations} $\phi_{\rm a} = \phi_{\rm r} = \phi$. 
The relation between these two quantities, $(S_{\rm a}/S_{\rm r})_{t^\prime}$
and $(S_{\rm a}/S_{\rm r})_{\phi}$, can be found taking into account that
in the observer's frame the ratio  of the  
times corresponding to equal angular separation of the sources from the
core is $(t_{\rm a}/t_{\rm r})_\phi = \mu_{\rm r}/\mu_{\rm a}$, while the
ratio of apparent times $t_{\rm a,r}=t^{\prime}/\delta_{\rm a,r}$  
corresponding to the same intrinsic time $t^{\prime}$ 
is $(t_{\rm a}/t_{\rm r})_{t^\prime} = \delta_{\rm r}/
\delta_{\rm a}$.
Using Eqs.~(2) and (3), we find that
$(t_{\rm a}/t_{\rm r})_\phi = \lambda (t_{\rm a}/t_{\rm r})_{t^\prime}\,$
where
\begin{equation}   
\lambda = \frac{\Gamma_{\rm r}\, \beta_{\rm r}\, \sin \theta_{\rm r}}
{\Gamma_{\rm a}\, \beta_{\rm a}\, \sin \theta_{\rm a}}\;\; \cdot
\end{equation}
Approximating the time evolution (decline) of the 
flare as $S_{\rm a,r} \propto t^{-p}$, as it was for the 19 March 
1994 flare of GRS 1915+105 (with $p\simeq 1.3$, Mirabel \& Rodriguez 1995), 
the relation between the flux ratios at equal intrinsic 
times and equal angular separations is found:
\begin{equation}
\left( \frac{S_{\rm a}}{S_{\rm r}}\right)_{t^\prime}\, = \, 
   \lambda^{ p} \, \left( 
\frac{S_{\rm a}}{S_{\rm r}}\right)_{\phi} \; \cdot
\end{equation}

For  GRS 1915+105 calculations  below result in
$\lambda \sim 0.85$, therefore the measured  
$(S_{\rm a}/S_{\rm r})_{\phi} \sim 8 $ corresponds to 
$(S_{\rm a}/S_{\rm r})_{t^\prime} \sim 6.5$. Further on we will omit 
subscript $(t^\prime)$, implying under $(S_{\rm a}/S_{\rm r})$ the flux ratio
at equal intrinsic times, if not mentioned otherwise.

Returning now to Eqs.~(2)--(4), note that for  identical jets, 
when  $\beta_{\rm a}=
\beta_{\rm r}=\beta $, $\theta_{\rm a}= \theta_{\rm r}=\theta$, 
Eq.(5) results in
$\lambda=1$, therefore in this particular case the equal angular separations
correspond to equal intrinsic times, as expected, so 
$S_{\rm a}/S_{\rm r} = (S_{\rm a}/S_{\rm r})_{\phi}$. Taking into 
account also that $L_{\rm a}^\prime =L_{\rm r}^\prime\,$, Eq.(4) comes to 
Eq.(1) with $k=3$. Then Eqs.~(1)--(3) make formally 
a system of 3 equations for 3 variables: $\beta$, $\theta$, $D$.
However, these equations are {\it not} independent when the power-law index 
$k$ is fixed, since they  {\it predict} the ratio 
$S_{\rm a}/S_{\rm r} = (\mu_{\rm a}/\mu_{\rm r})^{k+\alpha}$. Note that
due to high value of the index ${k+\alpha}\simeq 4$, the predicted flux ratio 
is very sensitive to variations of the angular velocities
$\mu_{\rm a,r}$ of the twin ejecta. 
In particular, 
in the case of $k=3$ and $\alpha = 0.84$, 
small error limits of the 
observed values of $\mu_{\rm a}=17\pm 0.4 \,\rm mas/d$ (milliarcsec/day) 
and $\mu_{\rm r}=9.0\pm 0.1 \,\rm mas/d$ (MR94) result in a rather broad range
of the flux ratios predicted for identical plasmoids,  
$(S_{\rm a}/S_{\rm r})_{\phi} = S_{\rm a}/S_{\rm r} = 13.1^{+1.8}_{-1.6}\:$, 
which is, nevertheless, beyond the observed  
$(S_{\rm a}/S_{\rm r})_{\phi} = 8\pm 1$.

However, this discrepancy disappears when an asymmetry 
between the ejecta is allowed.
Consider separately each of the 3 basic types of asymmetries
mentioned above.

\noindent
{\bf A}.~~$\beta_{\rm a}=\beta_{\rm r}=\beta \:$ {\it and} 
$\theta_{\rm a}=\theta_{\rm r}=\theta \:$, {\it but} 
$L_{\rm a}^\prime \neq L_{\rm r}^{\prime}$.~~ 
In this case the solution to Eqs.~(2) and (3)
for GRS 1915+105 is given in MR94:  
 $\beta \cos \theta =0.323$, and for the 
distance $D=12.5 \,\rm kpc$ the true speeds of the ejecta 
$\beta \approx 0.92$ and $\theta \approx 70^{\circ}$. Then from Eq.(4) follows
that in order to explain the  
ratio $(S_{\rm a}/S_{\rm r})_{\phi} \sim 8 $, the ratio of intrinsic 
luminosities
should be $L_{\rm r}^\prime / L_{\rm a}^{\prime}\sim 1.6$.

\noindent
{\bf B}.~~$\beta_{\rm a}=\beta_{\rm r}=\beta\:$ {\it and} 
$L_{\rm a}^\prime = L_{\rm r}^{\prime}\:$, {\it but}  
$\theta_{\rm a}\neq \theta_{\rm r}$.~~
Introducing dimensionless parameters 
$v_{\rm a}=\mu_{\rm a} D/c$ 
and $v_{\rm r}=\mu_{\rm r} D / c \,$, which correspond to the apparent speeds
(in units of speed of light) of the ejecta for a given distance $D$, 
the solution to Eqs.~(2)--(4) can be written as:
\begin{eqnarray} 
\tan \theta_{\rm a} & = & \frac{2 \, (s-1)\, v_{\rm a}}
{(s-1)^2 \,+ \, v_{\rm a}^2\,(b^2 \,- \,1) } \; ,\\
\sin \theta_{\rm r} & = & b \, \sin 
\theta_{\rm a} \;\; ,\\
\beta~~ & = & \frac{v_{\rm a}}{\sin \theta_{\rm a} \, + \,v_{\rm a}\,\cos 
\theta_{\rm a}} \;\;,
\end{eqnarray}
where $s \equiv (S_{\rm a}/S_{\rm r})^{1/(3+\alpha)}\,$, and 
$b\equiv \,s v_{\rm r}/v_{\rm a}= \,s \mu_{\rm r} / \mu_{\rm a}$. 
 Eqs.~(7)--(9) define 
$ \theta_{\rm a}, \, \theta_{\rm r}$  (in the range $\leq 180^\circ$) and 
$\beta$ for arbitrary $\mu_{\rm a}$, 
$\mu_{\rm r}$ and $s$ (note that if $b> 1$,  one should care for 
the condition $\sin\theta_{\rm r} \leq 1$).
However, starting from some distances these 
equations result in  $\beta > 1$, which becomes formally possible 
after reduction of $\Gamma_{\rm a}$ and $\Gamma_{\rm r}$ in Eq.(4).
For  the values of $\mu_{\rm a}$ and $\mu_{\rm r}$ 
observed for the March 19 event, and $S_{\rm a}/S_{\rm r} \leq 7$,  
the solutions with $\beta < 1$ require $D < 10.7 \,\rm kpc$. 
For example, assuming $D= 10\,\rm kpc$ and $S_{\rm a}/S_{\rm r} = 7$, 
Eqs.~(7)--(9) result in  
$\theta_{\rm a}=83.7^\circ$, $\theta_{\rm r}=57.7^\circ$ and $\beta= 0.92$.  
From Eq.(8) follows
that symmetrical solutions, with $\theta_{\rm a} = \theta_{\rm r}$, exist 
only if $b = 1$, which is not the case in GRS 1915+105. 

\noindent
{\bf C}.~~ 
$L_{\rm a}^\prime = L_{\rm r}^{\prime}\:$ {\it and} 
$\theta_{\rm a} = \theta_{\rm r} =
\theta \:$, 
{\it but} $\beta_{\rm a} \neq \beta_{\rm r}$.
In this  case,  due to very strong dependence of the flux 
ratio $S_{\rm a}/S_{\rm r}$ on the ratio  
$\Gamma_{\rm a}/\Gamma_{\rm r}$ (see Eq.(4)), the 
discrepancy between the observed
and `expected' flux ratios by factor of 
$(1.5-2)$ could be easily explained with only $\sim (10-20)\,\%$ difference 
in Lorentz factors of the oppositely moving plasmoids. 
Dividing Eq.(2) to Eq.(3), and substituting the resulting equation into
Eq.(4), we find the relation between $\beta_{\rm r} $ and $\beta_{\rm a} $,
and afterwards Eqs.~(2) and (3) can be resolved with respect
to $\beta_{\rm a}$ and $\chi \equiv \beta_{\rm a} \cos \theta$:
\begin{eqnarray}
\chi & = & \frac{(1 \, - \, b^2) \, v_{\rm a}^2 \, + \, s^2 \, -\, 1}
{(1 \, - \, b^2) \, v_{\rm a}^2 \, + \, s^2 \, +\, 2 s b \, + \, 1} \;\; , \\
\beta_{\rm a} & = & \sqrt{\chi^2 \,+ \, v_{\rm a}^2\, (1-\chi)^2} \;\;\; ,\\
\beta_{\rm r} & = & \beta_{\rm a} \frac{b}
{\sqrt{1\, +\, (b^2 -1)\,\beta_{\rm a}^2 } }\; \; .
\end{eqnarray}
Parameters $s$ and $b$ are the same as in Eqs.(7)--(9).
The angle $\theta$ is readily found after calculations of
$\chi$ and  $\beta_{\rm a}$.

In Fig.1 we show the dependence of the speeds of propagation,
$\beta_{\rm a}$ and $\beta_{\rm r}$, of the approaching and receding plasmoids 
 on the distance $D$, calculated from  
Eqs.~(10)--(12) for $\mu_{\rm a}=17\,\rm mas/d$, $\mu_{\rm r}=9\,\rm mas/d$,
and fixed flux ratio at equal intrinsic times 
$S_{\rm a}/S_{\rm r} = 7 $. For the distance $D=12.5\,\rm kpc$ the speeds 
$\beta_{\rm a}= 0.926 $ and $\beta_{\rm r}= 0.902$, and the angle $\theta
= 70.2^\circ$ are found. From Eq.(6) follows that in this case the flux ratio 
$S_{\rm a}/S_{\rm r} = 7 $ corresponds to 
$(S_{\rm a}/S_{\rm r})_{\phi} \simeq 8.6$.
Remarkably, the speeds of both ejecta are close to the 
value of $\beta=0.92$ 
found in MR94 assuming $\beta_{\rm a}= \beta_{\rm r}$. 
In Fig.2 we show the speeds $\beta_{\rm a}$ and 
$\beta_{\rm r}$ and relevant Lorentz factors $\Gamma_{\rm a}$
and $\Gamma_{\rm r}$ of counter ejecta, calculated for 
the same $\mu_{\rm a,r}$ and $\alpha$ as in Fig.1, but for different 
flux ratios in a broad range $5\leq S_{\rm a}/S_{\rm r} \leq 15 $. Note that  
{\it small} power-law index $1/(3+\alpha) = 0.26$ in  
the parameter $s=(S_{\rm a}/S_{\rm r})^{1/(3+\alpha)}$ significantly 
reduces the impact of 
uncertainties in the flux ratio $S_{\rm a}/S_{\rm r}$ 
on calculated speeds of propagations of the ejecta.

\section{Discussion}

The key assumption of this paper consists in allowance for an asymmetry 
between the
pair of plasmoids, which implies at least one of the 
enequalities: $L^{\prime}_{\rm a} \neq L^{\prime}_{\rm r}$, 
$\theta_{\rm a} \neq \theta_{\rm r}$, or $\beta_{\rm a} \neq \beta_{\rm r}$. 
Assuming that only
the angles of propagation 
are  different, $\theta_{\rm a} \neq \theta_{\rm r}$, one can explain the 
observed flux ratio 
$(S_{\rm a}/S_{\rm r})_{\phi} = 8\pm 1$ 
if GRS 1915+105 would be at a distance
$D\simeq 10 \,\rm kpc$, and requiring essential difference 
($\geq 25^\circ$) between $\theta_{\rm a} $ and $\theta_{\rm r} $. 
Note, however, that in this case the distance to GRS 1915+105 
does not agree with estimated kinematic distance
$D= 12.5 \pm 1.5 \: \rm kpc$ (MR94).
Besides, the radio observations of the March 19 outburst clearly 
show the motion of two sources in strictly opposite directions  
on the sky, within uncertainties of only few degrees (MR94).
Therefore the difference in 
the angles $\theta_{\rm a} $ and $\theta_{\rm r} $ could be only in the
plane of view.

More probable seem to be realization of the other two types of asymmetries. 
If the asymmetry is attributed mainly to 
different intrinsic (i.e. in the rest 
frames of the clouds) luminosities
$L^{\prime}_{\rm a} \neq L^{\prime}_{\rm r}$, then the receding 
cloud should be significantly ($\geq 50\,\%$) more luminous  than the 
approaching one. 
In principle,
such a difference in the luminosities of approaching and receding clouds
could be attributed to different  (by $\sim 30\,\%$) magnetic 
fields $B$, and/or to different amount of relativistic electrons in the  
ejecta.
At last, the observed flux ratio can be explained with only 
a small difference  in the speeds of propagation,  
$\beta_{\rm a} \neq \beta_{\rm r}$, corresponding to $\simeq 10\%$
difference between the Lorentz factors $\Gamma_{\rm a}$ and $\Gamma_{\rm r}$
of the bulk motion of the plasmoids. Due to very strong dependence of 
$S_{\rm a}/S_{\rm r}$  
on the ratio $\Gamma_{\rm a}/\Gamma_{\rm r}$,
a small asymmetry in the speeds of propagation of the jets can be 
considered as the main reason for the discrepancy between the measured and
`expected' flux ratios in GRS 1915+105. 
Note, however, that there is no particular reason to believe that 
only the velocities of the ejecta are different, and it is quite 
possible that the intrinsic luminosities of the clouds can be also
somewhat
different, $L_{\rm a}^\prime \sim L_{\rm r}^\prime$ 
(but $\theta_{\rm a} \approx \theta_{\rm r}$). 
In that case Eqs.~(10)--(12) can be used for calculations of 
the jet parameters, if one substitutes parameter 
$s=(S_{\rm a}/S_{\rm r})^{1/(3+\alpha)}$ there 
with $s_{\ast}=(S_{\rm a}\, L_{\rm r}^\prime/S_{\rm r}\, 
L_{\rm a}^\prime)^{1/(3+\alpha)}$. 

The asymmetry in the luminosities and/or 
speeds of propagation of the ejecta could be connected both with somewhat 
different conditions in the external medium at distances 
$\leq 10^{16}-10^{17}\,\rm cm$ from the core 
(reached by the ejecta in a few days)
or with an asymmetry of the bidirectional ejection  
itself. In the absence of a firmly established theoretical
model for development of the two-sided jets, 
there is no particular
reason to eliminate any of these two options. For example, different 
densities of the external medium might result in different speeds
of propagation, and/or different speeds of {\it expansion} of radio clouds  
leading to different magnetic fields in the clouds at equal intrinsic 
times. An asymmetry of the bidirectional ejection   
(which is obvious for
GRO J1655-44, see Hjellming \& Rupen 1995), e.g.  
$\sim 10\% $ difference in the masses 
or acceleration rates of the ejecta, could also result in   
different speeds of propagation and/or luminosities of the plasmoids.    

Remarkably, intrinsic asymmetry of bidirectional ejection process could
imply an 
interesting interpretation of the anticorrelation between the X-ray  and 
radio flares observed from
both microquasars GRS 1925+105 (Foster et al. 1996; Harmon et al. 1997) 
and GRO J1655-44 (Tingay et al. 1995; Harmon et al. 1995; Hjellming \& 
Rupen 1995),
which consists in appearance of strong radio outbursts generally after 
essential decline of the preceding X-ray flares, usually with some delay  
up to few days between these events.
Tingay et al. (1995) and Meier (1996)
interpreted this effect as an indication of suppression of jet production
mechanisms in the regime of supercritical accretion responsible for
strong X-ray flares, while ejection would occur only after slowing down of the
accretion rate to subcritical regime.
Possible asymmetry of the jets  allows us to suggest an alternative 
(in fact, to a large extent opposite)  
scenario where the jet production itself may result in temporary drop 
of the accretion from supercritical to subcritical regimes.

Indeed, internal energy in
the radio clouds, and correspondingly, minimum kinetic energy of the ejection
for strong outbursts in GRS 1915+105 is estimated as
$W \sim 10^{43} \rm erg$ (Liang \& Li 1995; Meier 1996), 
and even more (MR94).
Then the momentum transferred to 
relativistic ejecta can be estimated as $P_{\rm a,r} 
\sim W/c \geq 3\times 10^{32} \,\rm
g\,cm\,/ s$. Asymmetrical ejection generally would require, just from 
momentum conservation law, that significant {\it recoil} momentum 
$\Delta P = | {\bf P}_{\rm a} -
{\bf P}_{\rm r} |$ is transferred to the core of ejection, which is the third 
object in the interacting system "two jets + core". 
Since most probable site responsible for  
relativistic ejection is the inner part of the accretion disk (e.g., 
Blandford \& Payne 1982; Begelman et al. 1984), it is interesting to  
compare $\Delta P $ with the integrated value of the  
specific (per unit volume) momentum $\rho v = \rho | {\bf v}| $ of the gas 
orbiting in the inner accretion disk at radii $R^\prime \leq R$: 
$P_{\rm d}(R)=\int \rho  v  
{\rm d}^3 R^\prime$ (here 
$\rho $ is the mass density in the accretion flow). 
Using  the 
Shakura \& Sunyaev (1973) accretion disk model, this integrated `absolute
momentum' is estimated (in their {\it a}-region) 
as $P_{\rm d}\simeq 4\times 10^{21}\, 
\alpha_{\rm SS}^{-1}\,m^2 \, \dot{m}^{-1}\, r^3 \: \rm g\,cm\,/ s$, where
$\alpha_{\rm SS}\leq 1 $ is the viscosity parameter, and $r=R/R_{\rm g}$ is 
the radius in the units  
of $R_{\rm g} = 3\,(M/M_{\odot}) \,\rm km$.
 Then, even for $\alpha_{\rm SS}\sim 0.01$, the black hole mass   
$m=M/M_{\odot}\sim 10$, 
and accretion rate about the critical, $\dot{m}\sim 1$,  
a small recoil momentum $\Delta P \sim 0.1 \, {\bf P}_{\rm a, r}$  
being applied  to the jet production region (approximately, perpendicular to the
disk plane)
 would be enough  to cause significant 
changes, or even destruction, of the inner disk up to radii 
$r_{\ast} \sim 100$ estimated from condition 
$\Delta P \sim P_{\rm d}(r_{\ast})$, and 
resulting in a temporary reduction/termination of the fuel supply into this
region responsible for thermal X-rays. 

Thus, in the framework
of this scenario, the onset of subcritical/supercritical accretion  
would correspond to the active state, with X-ray flares which could 
proceed without significant ejection 
events and observable radio flares (see Foster et al. 1996; 
Tavani et al. 1996). 
Powerful ejection of 
radio emitting material may be accompanied by significant  destruction of the 
inner accretion zone, leading to a strong decline of the X-ray fluxes
simultaneously with production of relativistic ejecta. The ejection event, 
however, will be not accompanied by the {\it simultaneous} increase of the 
radio fluxes, since for an appreciable time the ejecta remain too compact
to be transparent at the radio frequencies. Thus, 
one has to expect a delay, up to a few days, between 
the instant of ejection and the appearance of strong radio flares, 
which time is needed for expansion of the plasmoids, with $v_{\rm exp}
\sim 0.1 \,c$,  to the radii $R_{\rm cl} \geq 5\times 10^{14}\,\rm cm$,  
as follows from the optical transparency of the radio plasmoids with respect to 
synchrotron self-absorption (Atoyan \& Aharonian 1997). 
Note that the radio clouds, moving with the speeds $v\sim c$, 
become transparent at distances 
$R \sim 10\, R_{\rm cl} \geq 5 \times 10^{15}\,\rm cm$, i.e. far away from 
the binary system. Depending on the timescale
needed for the accretion disk to 
recover after ejection event, 
the maximum of radio flares could appear just in the 
dips between X-ray flares (see Harmon et al. 1997). 
Rough estimate of the duration $\Delta t$ of such dips  as the dynamical time
of accretion from radii $r_{\ast}$ results in $\Delta t \propto 
r_{\ast}^{7/2}$, which could vary
on time scales from  minutes to days,  
depending on the spatial scales  
of the inner disk affected by ejection.

\vspace{3mm}
\noindent
{\bf Acknowledgments}~~~We thank H.J.V\"olk for valuable discussions. 
AMA acknowledges the support of his work 
through the Verbundforschung
Astronomie/Astrophysik of the German BMBF under the grant No. 05-2HD66A(7), 
and hospitality of the Max-Planck-Institute f\"ur Kernphysik during 
his visit to Heidelberg.

\vspace{5mm}


\centerline{\bf References}

\noindent{}{Atoyan, A. M., \& Aharonian, F. A. 1997, Proc. Workshop 
"Relativistic jets in AGN", Cracow, (in press)}

\noindent{}{Begelman, M. C., Blandford, R. D., \& Rees, M. J. 1984, 
Rev. Mod. Phys., 56, 255}

\noindent{}{Blandford, R. D., \& Payne, D. G. 1982, MNRAS, 199, 883}

\noindent{}{Bodo, G., \& Ghisellini, G. 1995, ApJ, 441, L69}

\noindent{}{Foster, R. S., et al. 1996, ApJ, 467, L81}

\noindent{}{Harmon, B. A., et al. 1995, Nature, 374, 704 }

\noindent{}{Harmon, B. A., et al. 1997, ApJ, 477, L85}

\noindent{}{Hjellming, R. M., \& Rupen, M. P. 1995, Nature, 375, 464}

\noindent{}{Liang, E., \& Li, H. 1995, A\&A, 298, L45}

\noindent{}{Lind, K. R., \& Blandford, R. D. 1985, ApJ, 295, 358}

\noindent{}{Meier, D. 1996, ApJ, 459, 185}

\noindent{}{Mirabel, I. F., \& Rodriguez, L. F. 1994, Nature, 371, 46}

\noindent{}{Mirabel, I. F., \& Rodriguez, L. F. 1995, 
Annals NY Academy of Science, 759, 21}

\noindent{}{Shakura, N. I., Sunyaev, R. A. 1973, A\&A, 24, 337 }

\noindent{}{Tavani, M., et al. 1996, ApJ, 473, L103}

\noindent{}{Tingay, S. J., et al. 1995, Nature, 374, 141}

\clearpage

\begin{figure}
\epsfxsize=15. cm
\epsffile[65 415 516 650]{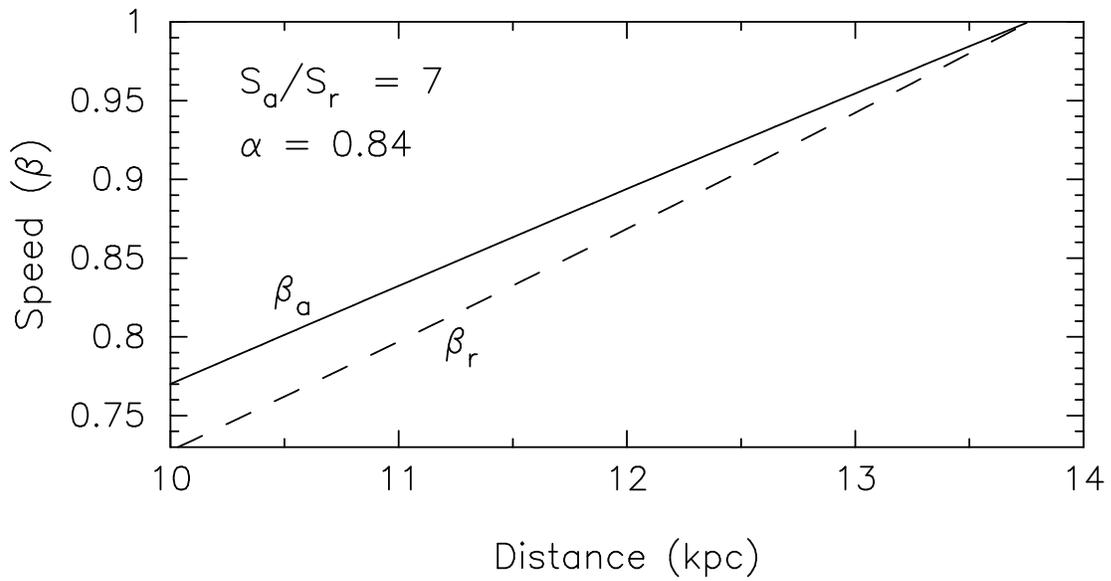}
\caption{The speeds of propagation of the approaching (solid line)
and receding (dashed line) asymmetrical pair of plasmoids calculated from 
Eqs.(10)-(12) assuming different distances to GRS 1915+105.}
\end{figure}

\begin{figure}
\epsfxsize=15. cm
\epsffile[65 210 513 552]{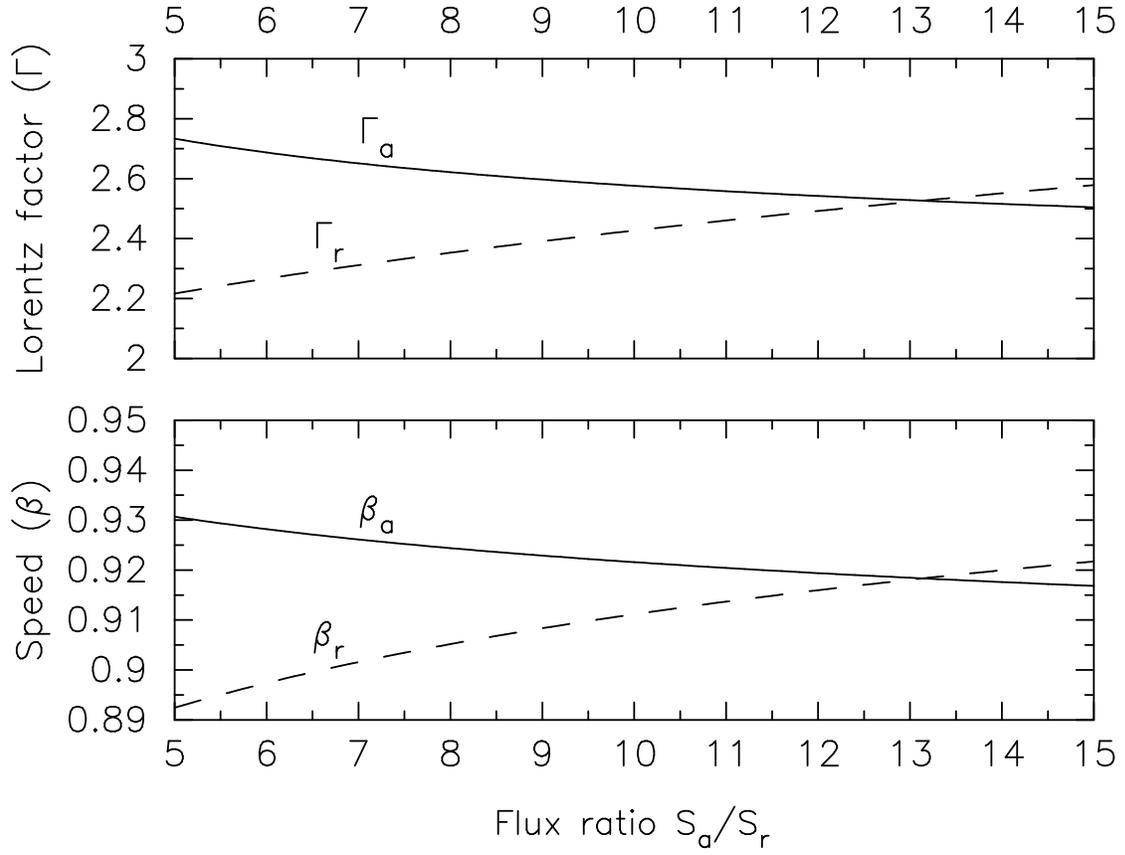}
\caption{The speeds (bottom panel) and the Lorentz factors 
(top panel) of the bulk motion of the pair of plasmoids calculated 
for a fixed distance $D= 12.5\,\rm kpc$ and angular 
velocities $\mu_{\rm a, r}$ observed from GRS 1915+105 (MR94), but assuming
different flux ratios $S_{\rm a}/S_{\rm r}$.}  
\end{figure}

\end{document}